\documentclass[aps,prc,superscriptaddress,showpacs,nofootinbib,floatfix,preprint]{revtex4}
\usepackage{graphicx}   
\usepackage{rotating}
\usepackage{dcolumn}
\usepackage{bm}
\usepackage{epsfig}
\usepackage{amssymb}
\usepackage{multirow}

\begin{document}

\title{Understanding the Role of Jet and Underlying Event in p+p and d+Au Collisions from PHENIX at RHIC}
\newcommand{\stonybrkc}{Chemistry Department, Stony Brook University, Stony Brook, NY 11794, USA}
\newcommand{\bnl}{Physics Department, Brookhaven National Laboratory, Upton, NY 11796, USA}
\affiliation{\stonybrkc}\affiliation{\bnl}
\author{Jiangyong Jia for the PHENIX Collaboration}
\date{\today}

%

\begin{abstract}
Dihadron azimuthal correlation measurements have revealed striking
modifications of the jets by the dense medium created in heavy-ion
collisions at RHIC. One important question is to what extent the
modification can be attributed to cold nuclear matter effects. In
this analysis, we carried out a detailed mapping of the correlation
patterns using high-statistics RUN8 d+Au minimum bias data. A
striking scaling behavior of the jet pair yields is observed at low
and intermediate $p_T$. The jet pair yields are found to be
enhanced relative to $N_{\rm coll}$ scaled p+p jet pair yields. The
nuclear modification factor for jet pair yields, $J_{\rm dAu}$,
seems to scale with $p_T^{\rm sum}$= $p_T^a+p_T^b$ (scaler sum),
and shows a characteristic Cronin-like enhancement at $p_T^{\rm
sum}<$5-7GeV/c. Interestingly, the level of yield modifications is
similar between the near- and away-side pairs, and the jet shapes
are not modified relative to p+p collisions. The pedestal yield
under the jet peak is studied in p+p and d+Au collisions. The
pedestal yield in p+p collisions is found to be larger than PYTHIA
calculations. In d+Au collisions, it is found to exceed a simple
sum of one p+p jet event and $\rm N_{coll}-1$ minimum bias p+p
events. The possible interpretation of these results and their
implications for Au+Au measurements are discussed.
\end{abstract}

\pacs{25.75.Dw 13.87.Fh}

\maketitle
\section{Introduction}
\label{intro}

It is generally believed that a strongly interacting Quark Gluon
Plasma (sQGP) is created in central Au+Au collisions at the
relativistic heavy ion collider (RHIC). The two most important
evidences for sQGP are the observation of a large elliptic flow and
a strong suppression of high $p_T$ jets (jet
quenching)~\cite{Adcox:2004mh,Adams:2005dq}. These results were
initially obtained from measurements of single particle
production~\cite{Adcox:2001jp,Adler:2003qi,Ackermann:2000tr}, and
were subsequently confirmed by various correlation
measurements~\cite{Adler:2002pu,Adler:2005ee}.

The information implied by correlation measurements, however, are
much richer than by single particle measurements. The complicated
correlation patterns can be better understood by studying them
separately in a high $p_T$ region and a low $p_T$ region. The high
$p_T$ region ($> 5$ GeV/$c$) is dominated by a suppressed but
essentially p+p like jet fragmentation component, which can be
interpreted as the combined result of jet quenching and surface
bias~\cite{Adare:2008cq}. The low $p_T$ region is characterized by
a highly non-trivial modification that depends on the $p_T$,
$\Delta\phi$ and $\Delta\eta$. This latter region seems to be
driven by a detailed balance between the jet and the medium: the
``jet'' signal exhibits the famous ``ridge'' and ''cone'' like
structures in $\Delta\eta$ and
$\Delta\phi$~\cite{Putschke:2007mi,Adare:2008cq}; the bulk medium
manifests a quark number scaling of elliptic
flow~\cite{Adare:2006ti} and an enhanced baryon/meson
ratio~\cite{Adler:2003cb}. Current efforts are focused on obtaining
a quantitative understanding of the mutual influences between the
jet and the medium~\cite{Jia:2008vk,Jia:2008kf}. Both the
modification of jets by the flowing medium and the response of the
medium to quenched jets need to be taken in to account.

\begin{figure}
\begin{center}
\epsfig{file=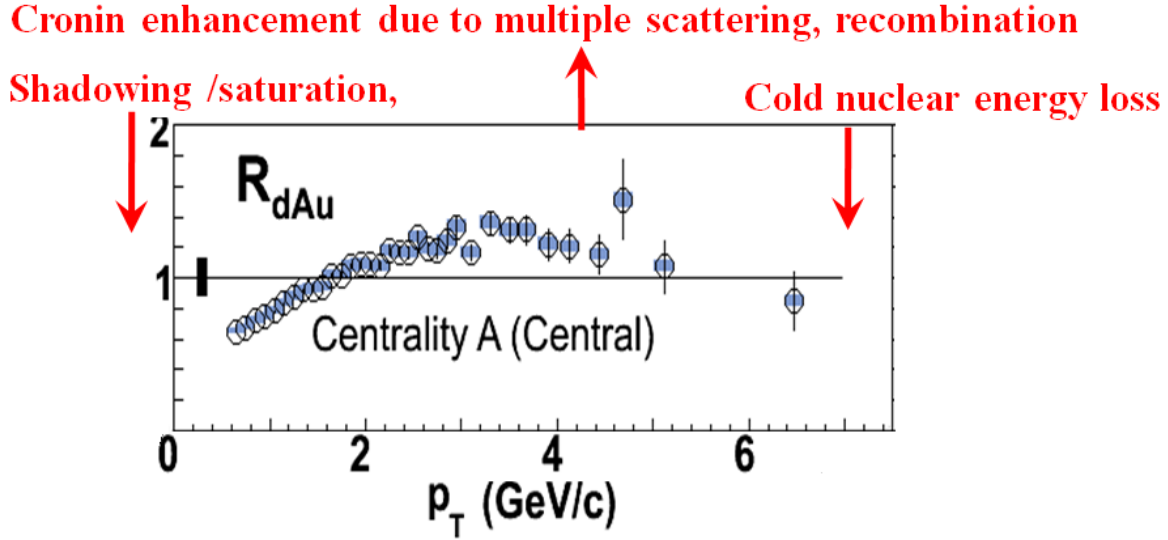,width=1.0\linewidth}
\caption{\label{fig:rda} Nuclear modification factor, $R_{dAu}$, for charged hadrons from RUN3 d+Au~\cite{Adler:2007by}.
Various cold nuclear effects responsible for the deviation from unity at different $p_T$ are indicated.}
\end{center}
\end{figure}

Study of single hadron production and dihadron correlation in p+A
or p+A-like collision such as d+Au provides important baselines for
understanding the results obtained in Au+Au collisions. Previous
measurements established the dominance of the final state effects
for the modifications of single particle production in Au+Au
collisions~\cite{Adler:2003ii}. However, various cold nuclear
effects such as initial parton distribution functions, Cronin
enhancements and cold nuclear energy loss etc, are also shown to be
not negligible~\cite{Adler:2006wg,Adler:2007by}. Their influences
depend on $p_T$ (see Figure~\ref{fig:rda}), i.e. shadowing effects
for low $p_T$ suppression, Cronin effects for intermediate $p_T$
enhancement, isospin and EMC effects for high $p_T$ suppression.
There is also some room for a cold nuclear matter energy loss,
which can reduce the overall yield~\cite{Vitev:2007ve}.

Dihadron correlation techniques have certain advantages over single
particle observables in constraining cold nuclear effects. While
single hadron spectra contain all contributions (including low
$Q^2$, non-perturbative processes), dihadron correlations are more
sensitive to hard-scattering processes. Previously published RUN3
d+Au results indicate little modification of jet
properties~\cite{Adler:2005ad}. However, those results are focused
primarily on a high $p_T$ region that might not be very sensitive
to cold nuclear effects. In this manuscript, we extend the
measurements to low and intermediate $p_T$, where the cold nuclear
effects are more pronounced. This study allows us to have a better
understanding of the final state effects in Au+Au collisions.

\begin{figure}
\begin{center}
\epsfig{file=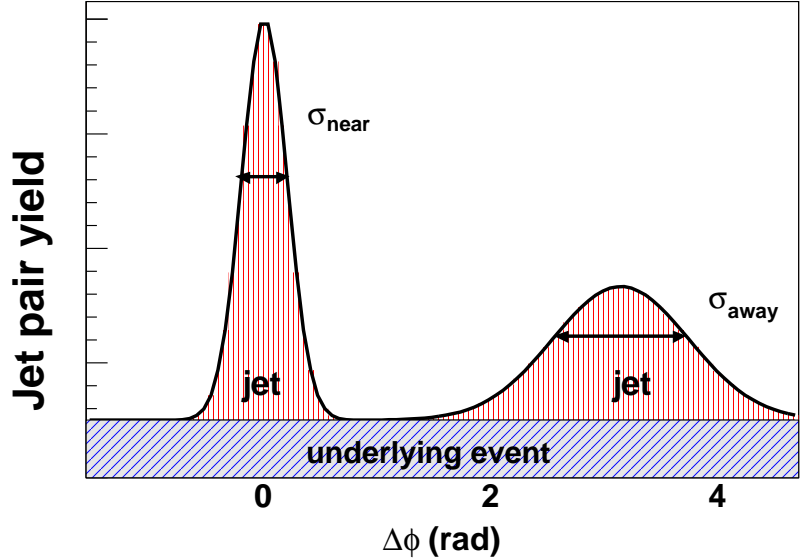,width=0.7\linewidth}
\caption{\label{fig:cartoon} Schematic illustration of dihadron azimuthal distribution
in for p + p and d + Au collisions. It has two peaks corresponding to
near- and away-side jet, and a component representing
the underlying event pairs.}
\end{center}
\end{figure}

Pairs in correlation analyses at RHIC energies are usually
decomposed into a jet part and underlying event (UE) part (see
Figure~\ref{fig:cartoon}). The jet part contains correlated pairs
from jet fragmentation and associated medium response, the UE part
includes the combinatoric pairs that are uncorrelated with the
trigger particle. Generally speaking, a rigorous decomposition is
already impossible in p+p collisions, due to non-perturbative, long
range correlations intrinsic to QCD. In A+A collisions, the
situation is further complicated by strong coupling between jet and
the flowing medium. The decomposition only makes sense if the
disturbance caused by the jet is localized; but if the disturbance
is dissipated to the whole medium, then all particles are
correlated with each other and the decomposition becomes very
difficult and highly model dependent. This problem is not unique to
dihadron correlations, it is also a serious issue for
multi-particle correlations and analyses requiring full jet
reconstruction.

Because of the intrinsic connection between jet and the UE, it is
important to study both of them simultaneously in heavy-ion
collisions, and to understand their mutual influences. Admittedly,
this is a difficult task. As a first step in addressing this
problem, we investigate the UE in simpler systems, i.e. p+p and
d+Au, which do not require elliptic flow subtraction. Our studies
not only provide the necessary handles on the cold nuclear effects,
but also fulfil longstanding interests of high energy physics
community~\cite{Affolder:2001xt,Acosta:2004wqa,Bahr:2008wk}. In p+p
collisions, the UE is one of the most important backgrounds for QCD
processes and for the Higgs boson search~\cite{Field:2006ek}. The
study of UE at RHIC energy can provide important inputs for tuning
the $\sqrt{s}$ dependence of phenomenological models, such as
PYTHIA~\cite{Sjostrand:2006za} and HERWIG~\cite{Bahr:2008pv}.

\section{Analysis}
This analysis uses the minimum bias p+p data from 2005 (2 billion
events) and d+Au data from 2008 run (1.6 billion events) at
$\sqrt{s_{\rm NN}} = 200$ GeV. A standard event mixing technique is
applied to correct for finite pair acceptance in azimuth, and a
ZYAM procedure~\cite{Adler:2005ee} is used to decompose the jet
function into a jet part and a $\Delta\phi$ independent pedestal.
Since the jet signal to background ratio is large in most kinematic
region considered here and jet width is rather narrow (except at
very low $p_T$), this should be a rather safe method for estimating
the jet yield.

In most jet correlation analyses, it is customary to use the
per-trigger yield, ${\rm PTY} = \frac{\rm PairYield}{\rm
TriggerYield}$, to measure the jet multiplicity. To quantify jet
modification, we usually compare jet yield with p+p collisions
using one of the three nuclear modification factors:
\begin{enumerate}
\item Modification of trigger yield \[R_{dAu}(p_T^a) =
    \frac{\rm TriggerYield_{dAu}}{\rm N_{coll}\times
    TriggerYield_{pp}}\]
\item Modification of pair yield \[J_{dAu}(p_T^a,p_T^b) =
    \frac{\rm PairYield_{dAu}}{\rm N_{coll}\times
    PairYield_{pp}}\]
\item Modification of per-trigger yield \[I_{dAu}(p_T^a,p_T^b)
    = \frac{\textrm{Per-Trigger Yield}_{\rm
    dAu}}{\textrm{Per-Trigger Yield}_{\rm pp}}\]
\end{enumerate}
where we use superscript ``a'' and ``b'' to indicate the two
particles in the pair. In the absence of nuclear effects, both pair
yield and trigger yield should scale with $\rm N_{coll}$, hence
deviations of $\rm J_{dAu}$ and $\rm R_{dAu}$ from unity can be
attributed to nuclear effects.

It is straightforward to show that the three quantities are related
to each other via the following relation:
\begin{eqnarray}
\label{eq:intro1}
\rm J_{dAu}\left(p_{T}^a,p_{T}^b\right) &=& \rm I_{dAu}\left(p_{T}^a,p_{T}^b\right) R_{dAu}\left(p_T^a\right)\\\nonumber &=&
\rm I_{dAu}\left(p_{T}^b,p_{T}^a\right) R_{dAu}\left(p_T^b\right)
\end{eqnarray}
This equation tell us that $I_{dAu}\neq1$ may not necessary imply
that jet is modified, it could be due to modification of trigger
yield. This could happen if some trigger particles do come from
jets and the yield of these triggers are modified. In this case,
the jet multiplicity is simply re-scaled by a constant factor
relative to p+p collisions, but the jet shape should remain
unchanged. For this reason, $\rm I_{dAu}$ may not be a good
variable for low $p_T$ correlation. Instead, we should use $\rm
J_{dAu}$, which directly reflects the jet yield modification.

\section{Jet Modifications in d+Au Collisions}

\begin{figure}
\begin{center}
\epsfig{file=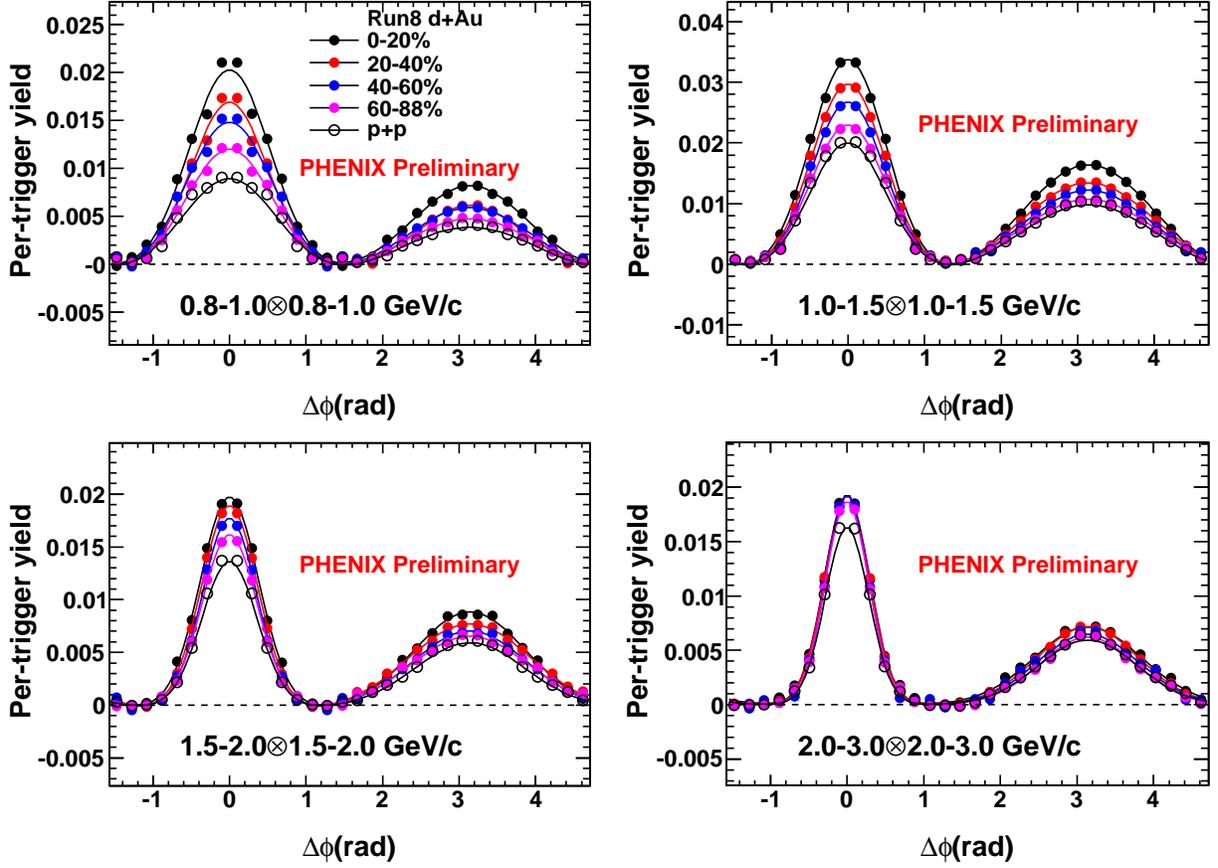,width=1\linewidth}
\caption{\label{fig:0p} Per-trigger yield distributions for fixed $p_T$ bins from low $p_T$ to high $p_T$. It demonstrates the disappearance of the modifications towards higher $p_T$.}
\end{center}
\end{figure}

\begin{figure}
\begin{center}
\epsfig{file=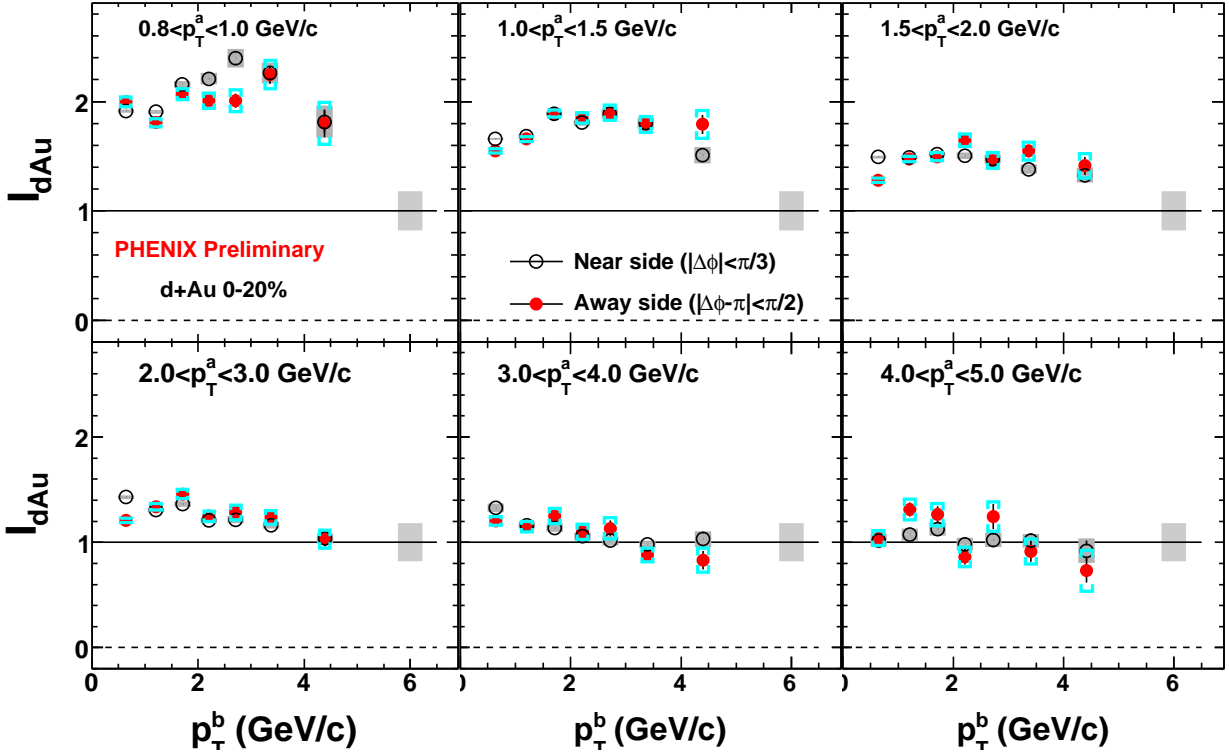,width=1.0\linewidth}
\caption{\label{fig:1p} $\rm I_{dAu}$ for the near-side and the away-side for 0-20\% centrality selection.}
\end{center}
\end{figure}

Modification of jet properties can be reflected by both the jet
shape and jet yield. Figure~\ref{fig:0p} shows a sample of
per-trigger yield distributions for p+p and d+Au collisions. A
clear enhancement of the amplitude is observed at both the
near-side and the away-side for d+Au collisions. The increase is
limited to the low $p_T$ region and disappears when trigger and
partner $p_T$ values rise above 2-3 GeV/$c$. This enhancement of
per-trigger yield is about factor of two at low $p_T$ as measured
by $\rm I_{dAu}$ in Figure~\ref{fig:1p}). It gradually disappears
at large trigger and partner $p_T$.

\begin{figure}
\begin{center}
\epsfig{file=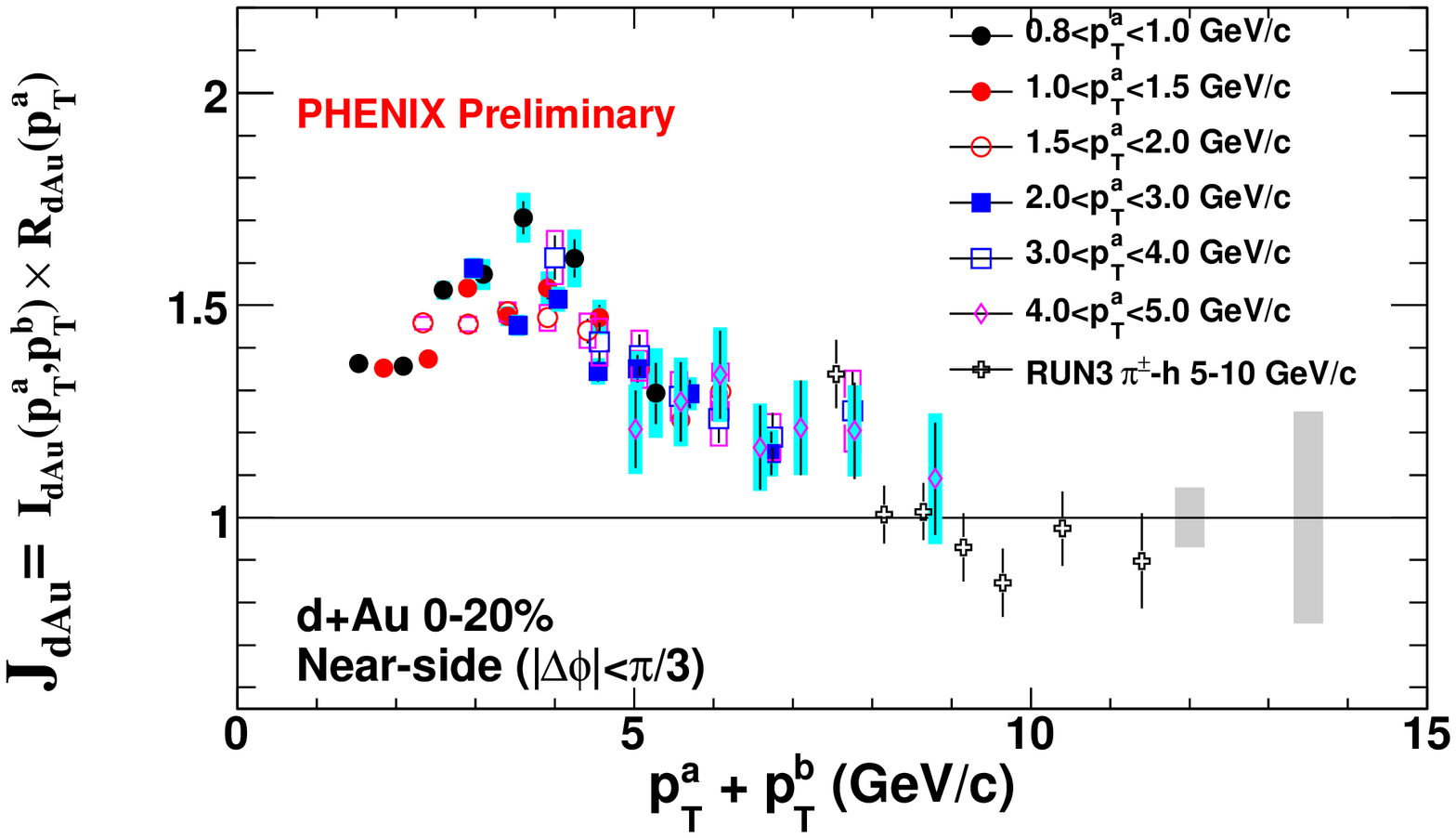,width=0.9\linewidth}
\epsfig{file=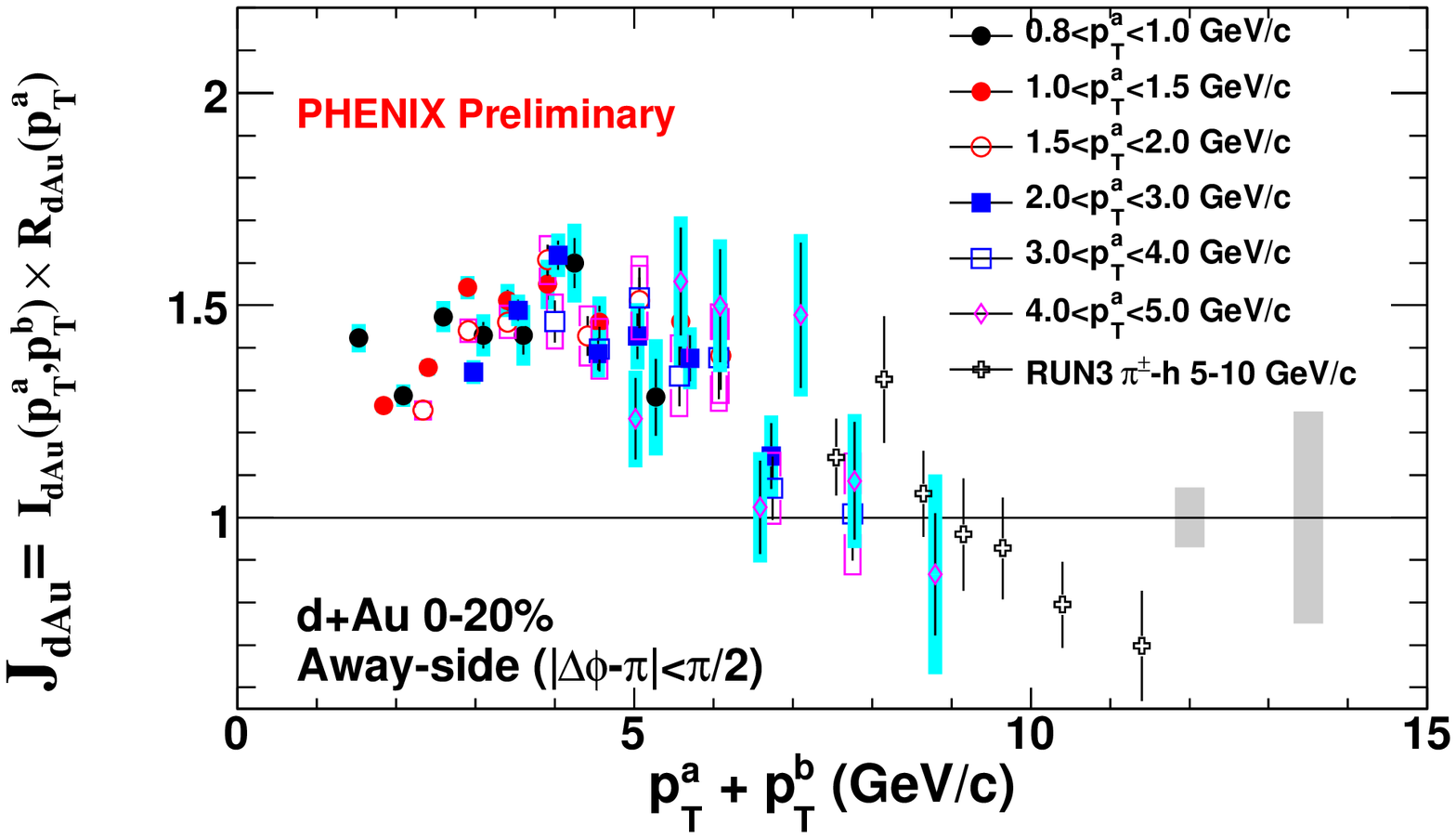,width=0.9\linewidth}
\caption{\label{fig:2p} $\rm J_{dAu}$ in 0-20\% d+Au centrality selection versus $\rm p_{T}^{sum}=p_T^a+p_T^b$ for near-side (top panel) and away-side (bottom panel).}
\end{center}
\end{figure}

\begin{figure}
\begin{center}
\epsfig{file=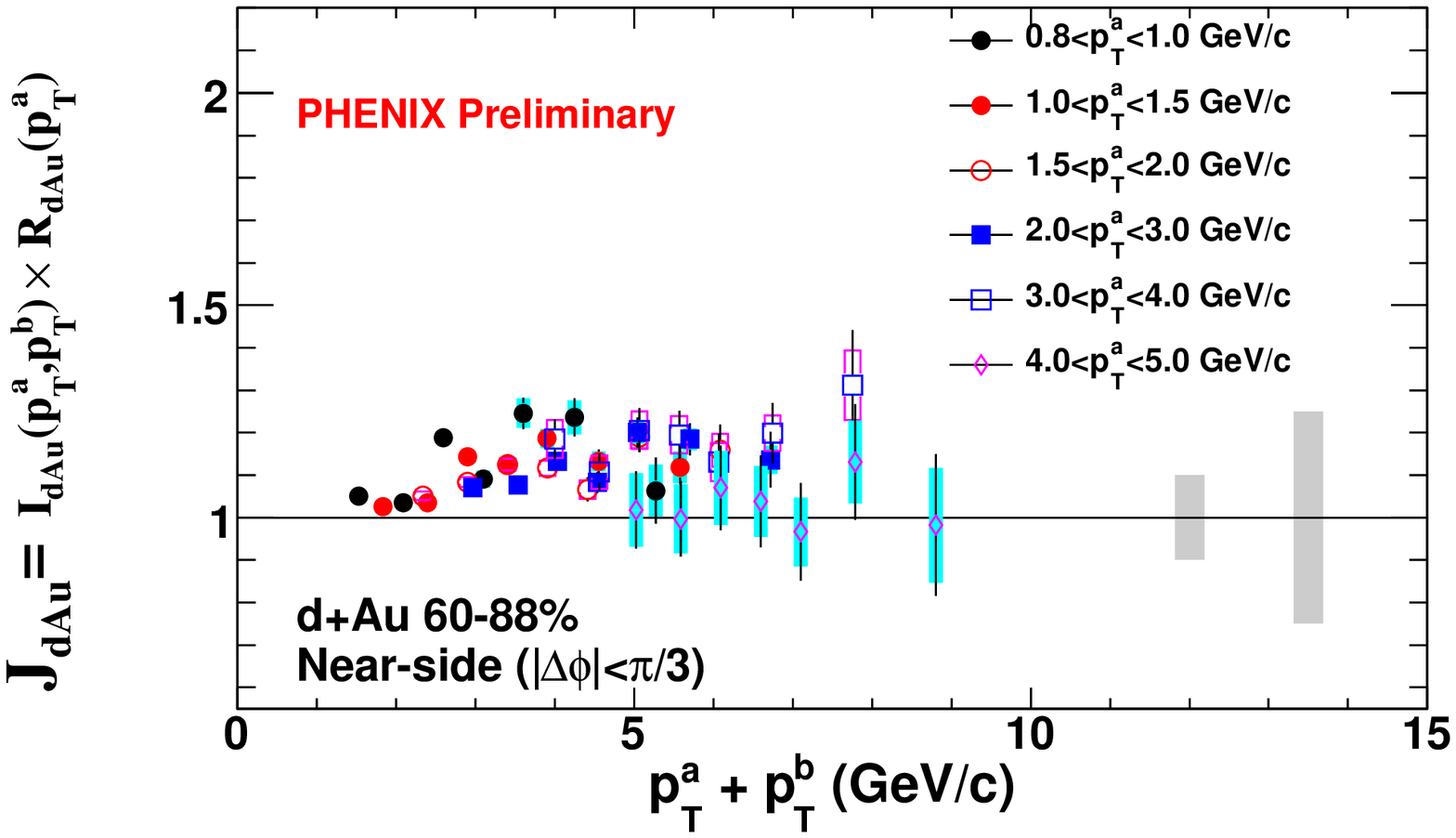,width=0.49\linewidth}
\epsfig{file=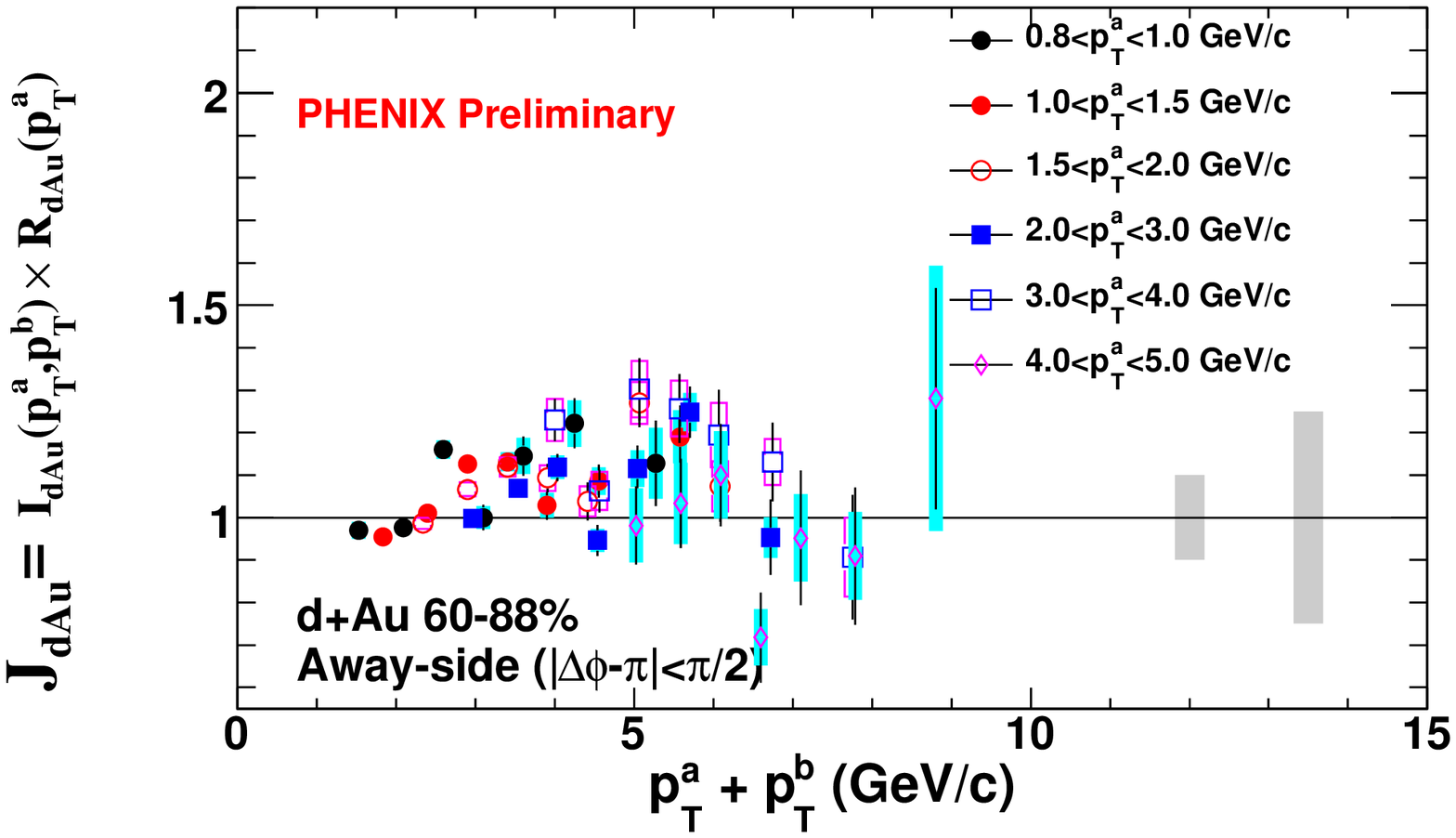,width=0.49\linewidth}
\caption{\label{fig:3p} $\rm J_{dAu}$ in 60-88\% d+Au centrality selection versus $\rm p_{T}^{sum}=p_T^a+p_T^b$ for near-side (left panel) and away-side (right panel).}
\end{center}
\end{figure}

As we argued before, a fraction of the enhancement seen in
Fig.~\ref{fig:1p} is due to the suppression of trigger yield at low
$p_T$ (see Fig.~\ref{fig:rda}, which increases the per-trigger
yield. To avoid that, we measure instead the absolute jet pair
yield per event and construct the pair yield nuclear modification
factor, $\rm J_{dAu}$, for each combination $p_T^a$ and $p_T^b$. A
compilation of $\rm J_{dAu}$ measurements are shown in
Figure~\ref{fig:2p} for central d+Au collisions. It is plotted as a
function of pair proxy energy $\rm p_{T}^{sum}=p_T^a+p_T^b$,
separately for the near- and the away-side. The high $p_T$
$\pi^{\pm}-h$ correlation data from RUN3 d+Au
collisions~\cite{Adler:2005ad} are included in this compilation.
Figure~\ref{fig:2p} shows that the $\rm J_{dAu}$ values
approximately follow a common curve. It increases with $\rm
p_{T}^{sum}$ at low $p_T$, and peaks at a level significantly above
one around $\rm p_{T}^{sum}\approx$4 GeV/$c$, then decreases
towards larger $\rm p_T^{sum}$. We do not see much yield
modification for peripheral d+Au collisions (Figure~\ref{fig:3p}).
The shape of the enhancement of Figure~\ref{fig:2p} resembles the
Cronin-like peak seen in single particle production. However, the
level of enhancement is much bigger and it does not exhibit
shadowing-like suppression at low $p_T$.

\begin{figure}
\begin{center}
\epsfig{file=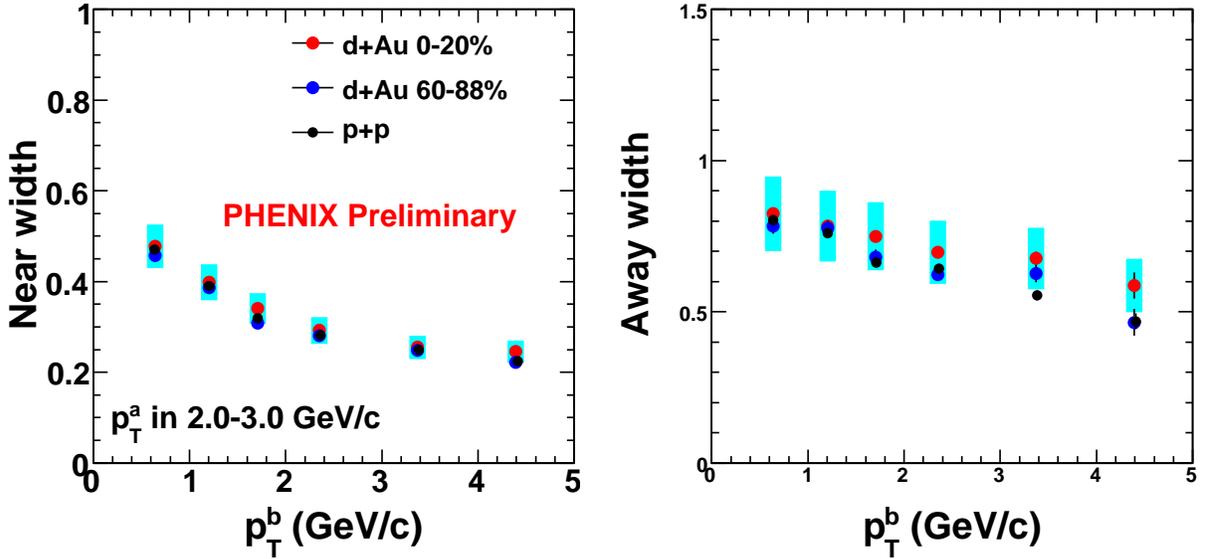,width=1\linewidth}
\caption{\label{fig:0pb} Near-side width (left panel) and away-side width (right panel) obtained from double Gaussian fit in central d+Au (0-20\%), peripheral d+Au( 60-88\%) and $p+p$ collisions.}
\end{center}
\end{figure}

To study the modification of jet shape, we fit the jet yield
distribution with a double Gaussian function and extract the near-
and away-side width. Some examples of such fit are shown in
Figure~\ref{fig:0p}. The summary of the Gaussian widths from the
fit are shown in Figure~\ref{fig:0pb}. The near-side widths are
identical between p+p and d+Au. The away-side widths indicate a
small broadening in central d+Au collisions, which is expected from
multiple scattering effects. But the level of broadening is well
within the quoted 15\% systematic errors. Work is underway to
refine the systematic errors, such that we can better quantify this
broadening in the future.

\section{The Underlying Event in p+p and d+Au Collisions}

\begin{figure}
\begin{center}
\epsfig{file=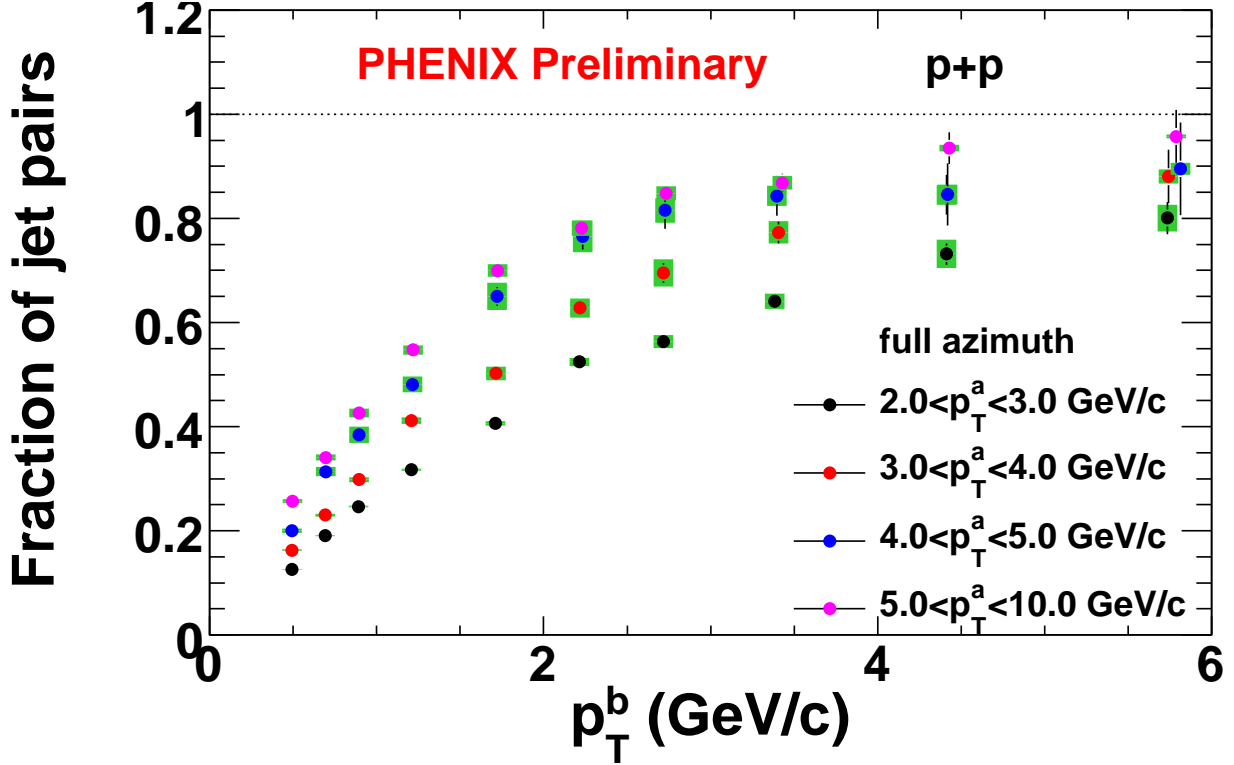,width=1\linewidth}
\caption{\label{fig:10p} Fraction of pairs in p+p collisions contained in the jet peak, integrated over the full 2$\pi$ range.
The systematic errors shown reflect the ZYAM uncertainties.}
\end{center}
\end{figure}

Having examined the jet shape and yield modifications, we proceed
to study the properties of the underlying event (UE). A natural
question to ask is what is the relative contribution from the jet
and the UE. The answer to this question gives us a first order
estimation of the contribution from hard-scattering processes in
p+p and d+Au collisions. And it is essential for us to properly
understand the modification of these processes in Au+Au collisions.
Figure~\ref{fig:10p} shows the fraction of pairs in the jet peak
(jet pair fraction) as a function of $p_T$ in p+p collisions. The
jet pair fraction is already more than 15\% at lowest $p_T$ bin
($p_T^a\approx2$ and $p_T^b\approx0.5$ GeV/$c$) and quickly
increases to close to 100\% at high $p_T$. This is much bigger than
the typical level (few percent) seen in central Au+Au collisions.
We want to point out that this estimation only gives the lower
limit of jet contribution, since the UE contains contributions from
uncorrelated jets as well as those from large angle radiations.

In heavy ion collisions, the jet signal is difficult to extract
because of the large UE level. In p+p and d+Au collisions, we face
the opposite problem: jet signal is so big that the UE can be
strongly influenced by effects associated with the hard-scattering
process such as the initial and final state radiation. Our
definition of the UE is different from previous approaches used by
the CDF Collaboration~\cite{Affolder:2001xt}, where the azimuth
correlation is made between reconstructed jets and charged hadrons.
This method is cleaner in separating the jet from the UE. However,
the dihadron correlation method is still useful for two reasons.
First, full jet reconstruction is problematic and questionable at
$p_T<10$ GeV/$c$, where dihadron correlation method can still be
used. Second, it is currently the only method which allows a
systematic study in p+p, d+Au and Au+Au at RHIC. In order to avoid
potential confusion, we thereafter refer the UE obtained by
dihadron correlation method as ``pedestal''.

Due to the need for event mixing to account for finite azimuthal
acceptance, PHENIX usually defines the correlation function as the
ratio of pair distributions from same-events and
mixed-events~\cite{Adare:2008cq}, where each is normalized
separately by the number of events:
\begin{eqnarray}
\label{eq:cf}
C(\Delta\phi)=\frac{N_{fg}(\Delta\phi)}{N_{mix}(\Delta\phi)}=\frac{\rm Jet_{pairs}+
UE_{pairs}}{\left<n_a\right>\left<n_b\right>}
\end{eqnarray}
In the second part of the equation, we decompose the foreground
into the jet part and pedestal part via the ZYAM approach, and we
use the fact that mixed event yield equal to the product of trigger
yield, $\left<n_a\right>$, and partner yield, $\left<n_b\right>$.
Because this way of constructing the correlation function, it is
convenient for PHENIX to measure the pedestal yield relative to the
p+p single particle yield. We define a ratio $\zeta$:
\begin{eqnarray}
\label{eq:i}
\zeta = \frac{\rm UE_{pairs}/\left<n_a\right>}{\left<n_b\right>}=\frac{\textrm{Assoc. Pedestal Yield Per-trig}}{\textrm{Min. Bias yield Per-event}}
\end{eqnarray}
The advantage of this quantity is that the tracking efficiency
cancels in the ratio. One can simply multiply $\zeta$ with
published single particle yield~\cite{Adler:2005in} to obtain the
pedestal yield.

\begin{figure}
\begin{center}
\epsfig{file=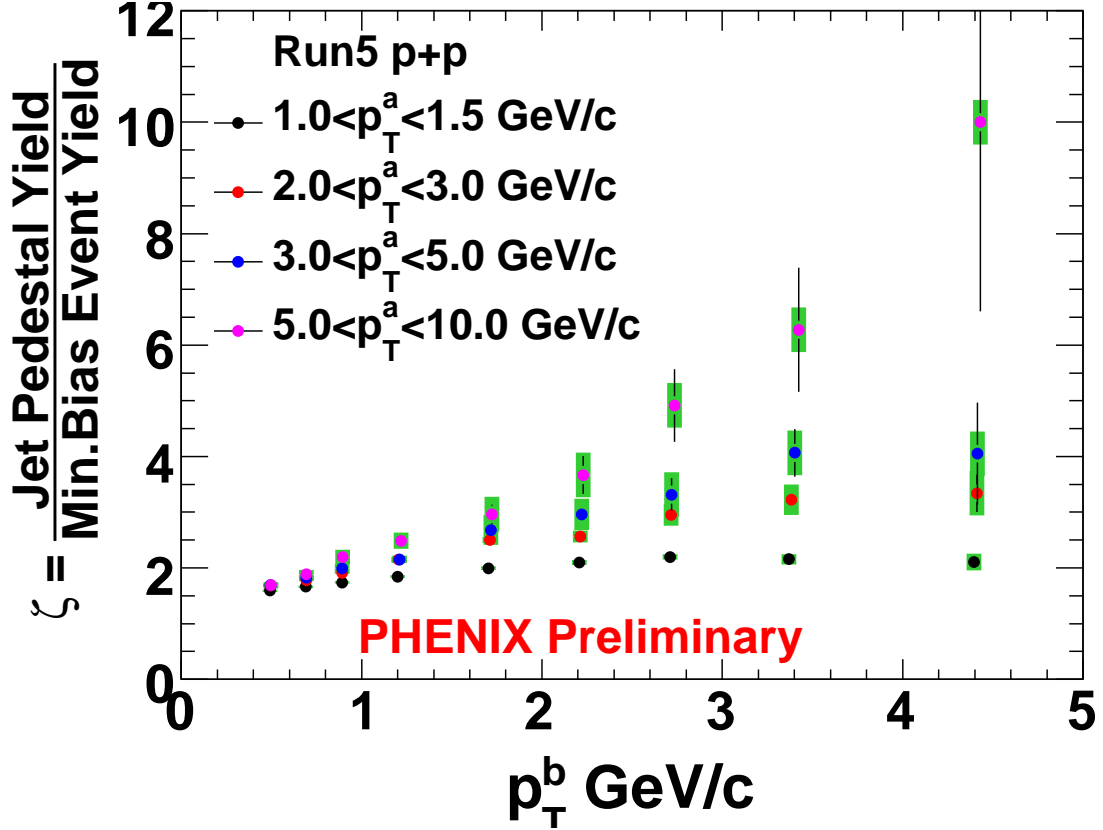,width=0.9\linewidth}
\caption{\label{fig:11p} The summary of the $\zeta$ in various trigger $p_T$ bins as function of associated $p_T$ in p+p collisions.}
\end{center}
\end{figure}

Figure~\ref{fig:11p} summarizes $\zeta$ values for various trigger
$p_T$ bins as a function of partner $p_T$. The $\zeta$ values are
always above one and increase strongly with the $p_T$ of the two
hadrons. The change with $p_T$ may be due to increase of the
initial and final state radiation. The fact that $\zeta>1$ can be
attributed to a centrality bias in p+p events, caused by the
requirement of a high $p_T$ hadron pair. The impact parameter for
these events are usually smaller than that for the minimum bias
events used for the mixing, which may lead to more underlying event
activity due to multiple parton-parton interactions (MPI).

\begin{figure}
\begin{center}
\epsfig{file=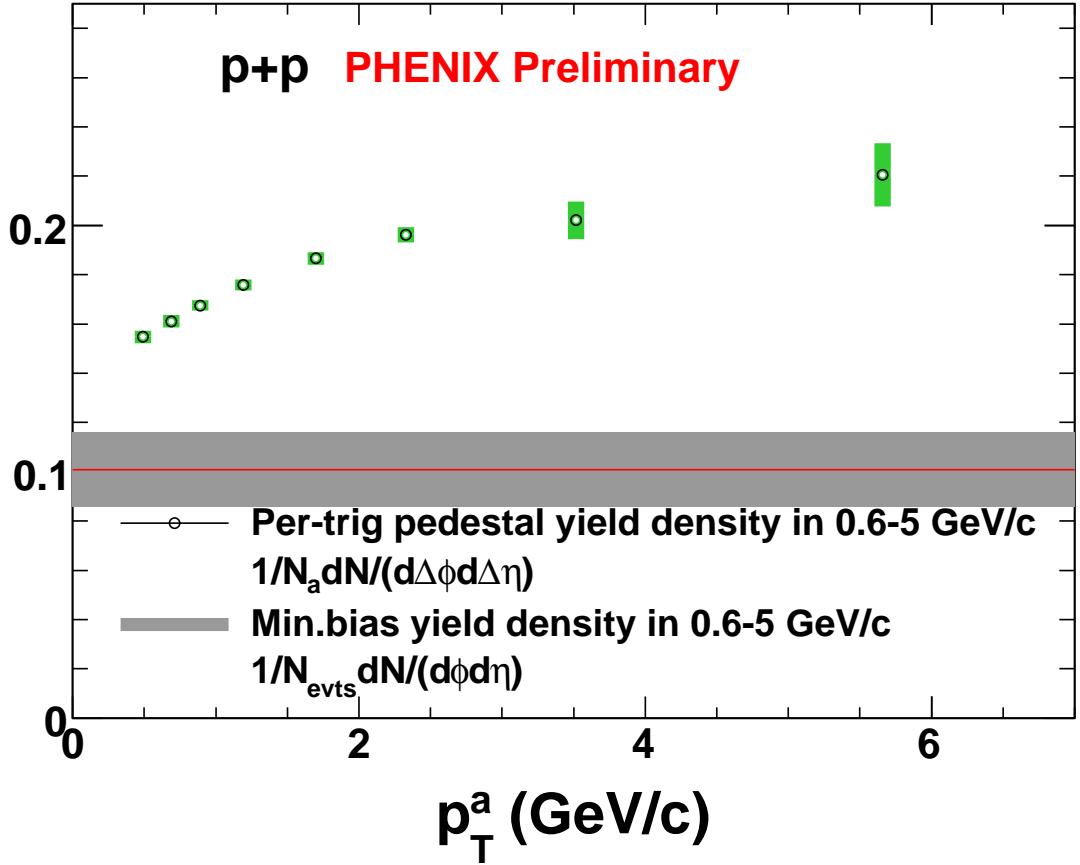,width=1.0\linewidth}
\caption{\label{fig:13pb} Pedestal yield per-trigger per unit of azimuth and pseudo-rapidity as a function of trigger $p_T$.
The shaded band indicates the yield from minimum bias p+p yield (corresponds to 21.9mb).}
\end{center}
\end{figure}

Figure~\ref{fig:13pb} compares the pedestal yield per-trigger
integrated from $0.6<p_T^b<5$ GeV/c as a function of trigger $p_T$.
The pedestal yield is expressed as transverse density per unit of
azimuth and pseudo-rapidity, $1/N_{a}dN/d(\Delta\eta\Delta\phi)$,
similar to~\cite{Affolder:2001xt}. The integrated UE yield
increases rapidly at low trigger $p_T$, then increases more slowly
at high trigger $p_T$. Again, the initial increase is related to
the increase of MPI, while the slower rise for larger trigger $p_T$
is mainly due to the initial and final state radiation effects.

Because of its non-perturbative nature, our current understanding
of the UE is mainly obtained through tuning phenomenological monte
carlo event generators, such as PYTHIA and
JIMMY~\cite{Butterworth:1996zw}, to match the data. In this
analysis, we compare our results with calculations from PYTHIA.
PYTHIA has a large set of parameters that can affect the jet yield
and the UE yield. To facilitate the usage, PYTHIA pre-packages
several collections of default parameter values, known as PYTHIA
tunes (since version 6.410), each is identified with a unique
integer number~\cite{pythia}. One of the popular tunes is the Rick
Field tune that can reproduce the CDF RUN2 data, known as TUNE A
(100). Another popular tune is TUNE S0A, which is based on the new
PYTHIA UE framework introduced since version 6.3. The comparison
also depends on the jet fragmentation scheme because we do the
azimuthal correlation at the hadron level instead of reconstructing
the full jet. In this analysis, we considered the following three
different settings from PYTHIA 6.419.
\begin{itemize}
\item TUNE A (100), with parameter set tuned to CDF by Rick
    Field, string fragmentation.
\item TUNE A (100), but with independent fragmentation.

\item TUNE S0A(303), with New UE/MI framework, string
    fragmentation.
\end{itemize}

\begin{figure}
\begin{center}
\epsfig{file=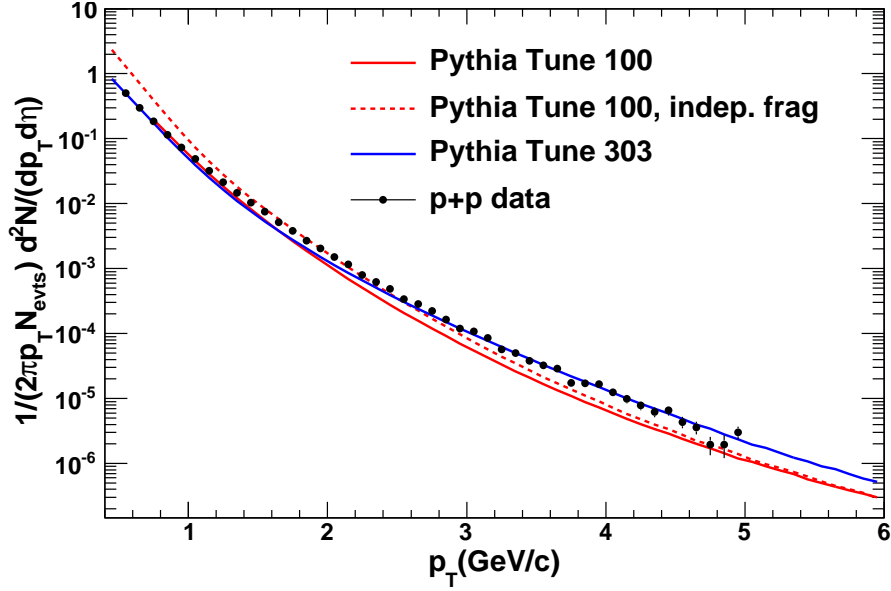,width=0.75\linewidth}
\caption{\label{fig:14} Comparison of PYTHIA tunes with the charged hadron spectra.}
\end{center}
\end{figure}

\begin{figure}
\begin{center}
\epsfig{file=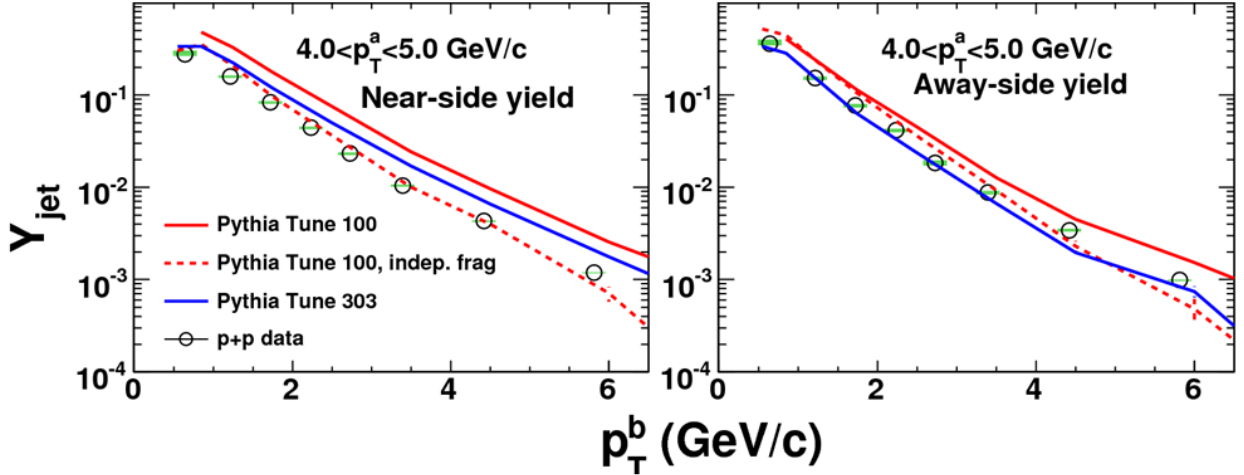,width=1\linewidth}
\caption{\label{fig:b15} Comparison of PYTHIA tunes with the per-trigger yield at the near-side (left panel) and away-side (right panel).}
\end{center}
\end{figure}

\begin{figure}
\begin{center}
\epsfig{file=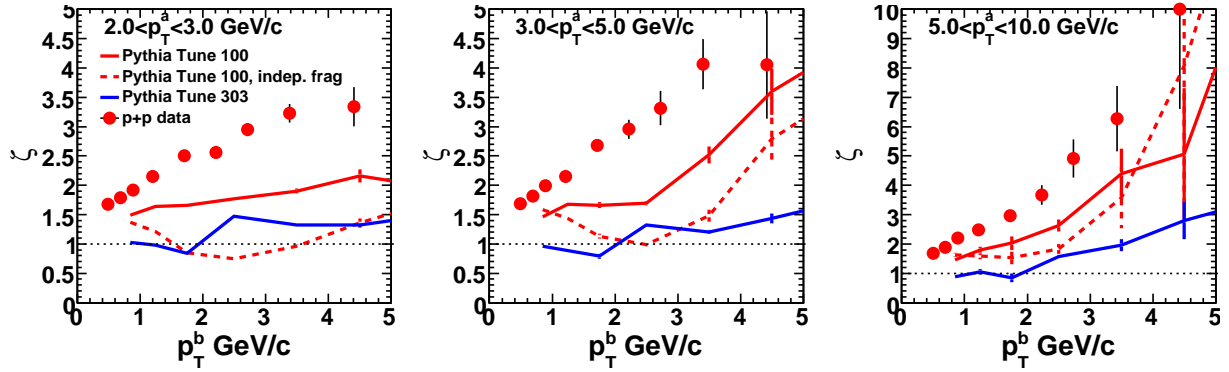,width=1.0\linewidth}
\caption{\label{fig:16} Comparison of the $\zeta$ values with
PYTHIA tunes in three trigger $p_T$ ranges.}
\end{center}
\end{figure}

For each setting, we generate enough simulated events, repeat the
same ZYAM procedure as for the real data analysis, extract the jet
yield and pedestal yield and compare with our measurements. A
satisfying parameter setting should be able to reproduce all
observables : single hadron spectra, jet yield and pedestal.
Figure~\ref{fig:14}-\ref{fig:16} show such comparison of the data
to all three settings. None of the settings can describe all three
observables simultaneously. In general, TUNE A with string
fragmentation over-predicts the jet yield, but has the best match
for the pedestal yield; TUNE A with independent fragmentation does
a better job for the jet yield but it does a equally poor job for
single spectra; TUNE S0A with string fragmentation can describe the
single spectra and jet yield reasonably well, but greatly
under-predicts the pedestal yield~\footnote{We would like to point
out our UE study focuses at low $p_T$ ($<5$ GeV/c), it has been
shown that the UE at high $p_T$ in general has better agreement
with the PYTHIA tunes~\cite{star}}. These comparisons show that our
data can be used to optimize the parameter values at RHIC energy,
and to tune their dependencies on collision energy.

\begin{figure}
\begin{center}
\epsfig{file=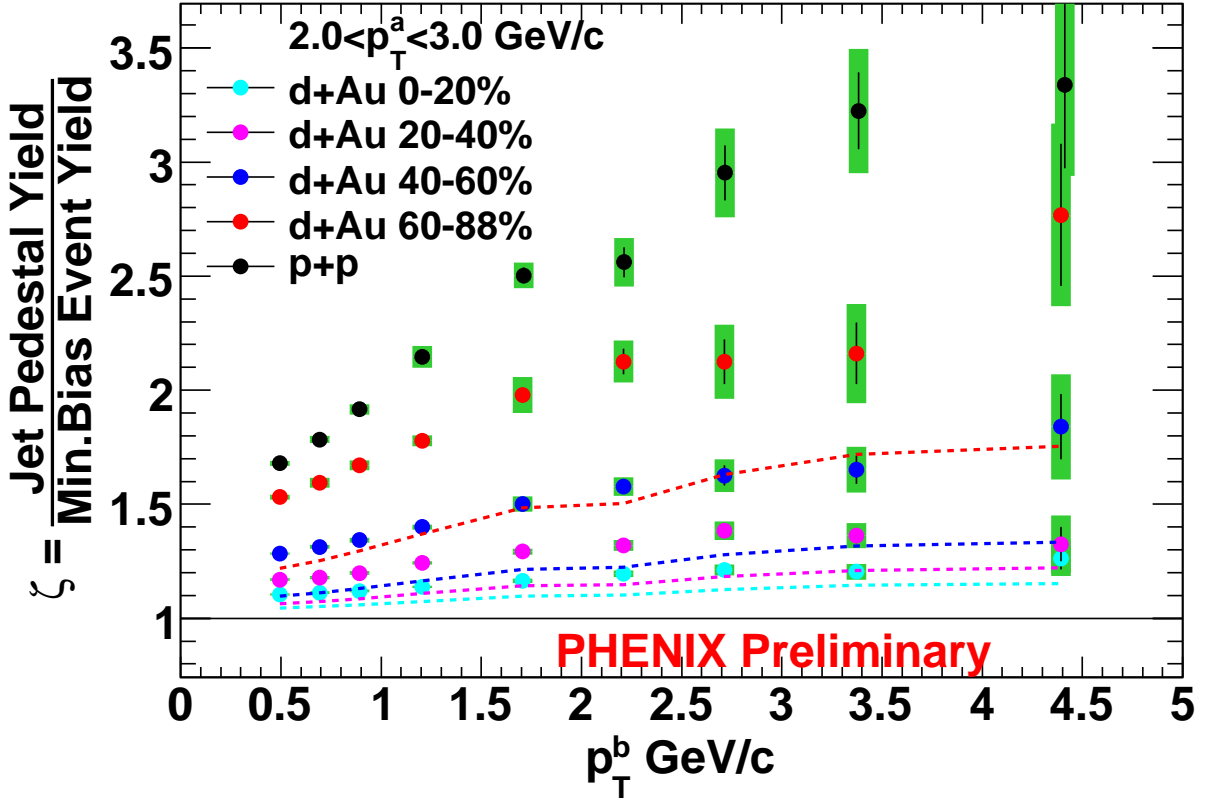,width=1\linewidth}
\caption{\label{fig:14p} The Centrality dependence of the $\zeta$ in d+Au collision and compared with $p+p$. The dashed lines are the estimated pedestal level based on Eq.~\ref{eq:4}.
From top to bottom, the four lines corresponds to 60-88\%, 40-60\%, 20-40\% and 0-20\% d+Au collisions, respectively}
\end{center}
\end{figure}

If the enhancement of the pedestal is due to MPI (whose probability
is proportional to overlap function), we should expect a strong
centrality dependence of the pedestal yield in d+Au collisions. A
systematic study of the pedestal yield from p+p to d+Au collisions
thus can help us to understand the nature of the MPI in p+p.
Figure~\ref{fig:14p} shows the $p_T$ dependence of $\zeta$ values
for p+p and d+Au collisions. The $\zeta$ value at fixed $p_T$
decreases towards central d+Au collisions. The central d+Au
collisions also show much weaker increase of $\zeta$ value with
$p_T$. These results suggest that the centrality bias due to the
triggering condition is much smaller in d+Au collisions than in
p+p. In this case, the pedestal yield should be very close to the
single particle yield for the corresponding centrality bin.

A simple model is used to understand the scaling behavior of the
pedestal yield. In this model, we assume that hard-scattering
giving rise to the jet signal occurs only in one nucleon-nucleon
collision in each foreground d+Au event, all other nucleon-nucleon
collisions in the same event are assumed to be the same as the
normal minimum bias p+p collision. In this case, the ambient
particle production should scale as $R_{dAu}$. The pedestal yields
in p + p and d + Au, $\rm UE_{dAu}$ and $\rm UE_{pp}$, should be
related to each other through the following relation.
\begin{eqnarray}
\label{eq:3}
{\rm UE_{dAu} = UE_{pp} + R_{dAu}(N_{coll}-1) Min.BiasYield_{pp}}
\end{eqnarray}
From this equation, we can derive the following relation relating
the $\zeta$ values in d+Au and p+p
\begin{eqnarray}
\label{eq:4}
{\rm \zeta_{dAu}N_{coll} = \zeta_{pp}/\epsilon_{mb} +  R_{dAu}(N_{coll}-1)}
\end{eqnarray}
where $\epsilon_{\rm mb}$=0.694 is the correction factor accounting
for p+p trigger efficiency in PHENIX. The calculated $\zeta$ values
for d+Au collisions from this formula are indicated by the curves
in Figure~\ref{fig:14p}. Our model under-predicts the data at high
$p_T$ and in central d+Au collisions, suggesting that more
activities for ambient nucleon-nucleon collisions are preferred. It
may also imply that initial state radiation associate with
hard-scattering can further scatter with the Au nuclei, thus
further increase the pedestal yield.

\section{Discussion}

The results shown in Figure~\ref{fig:2p}-\ref{fig:0pb} are
intriguing: while the jet shape in d+Au collisions is not modified
much, the yield of correlated pairs is significantly enhanced
relative to binary scaled p+p collisions. This observation seems to
imply that hard-scattering cross section is enhanced but jet
fragmentation and dijet acoplanarity are not modified much.
Furthermore, the underlying event level in d+Au collisions exceeds
their corresponding single particle yield (Figure\ref{fig:14p}).
This means that the cold nuclear matter effects already contribute
a significant fraction of yield enhancement seen in Au+Au, but the
shape modification in Au+Au collisions is mostly due to final state
effects in dense medium. To make sense of these results, we examine
in the following the list of known effects in d+Au collisions, and
speculate the roles they may play in dihadron correlation results.
\begin{itemize}
\item Shadowing/CGC effect: This effect is responsible for the
    suppression of the single particle yield at low $p_T$. But
    our data seems to imply that it does not suppress the
    hard-scattering processes in the kinematic range considered
    in the analysis.

\item Power corrections~\cite{Qiu:2003vd}: The expectation of
    binary scaling relies on the factorization theorem, which
    should break down at small $Q^2$. The lowest pair momentum
    in our analysis is at around 1-1.5 GeV/$c$, which
    corresponds to a $Q^2$ value starting at around
    $Q^2\approx1-2$(GeV/$c$)$^2$. At such small $Q^2$ value,
    the modification jet signal due to power corrections might
    be significant.

\item Multi-parton
    collisions~\cite{Strikman:2001gz,Frankfurt:2004kn}: This
    refer to the situation where more than one hard-scattering
    happens in a given event, such processes in general break
    the factorization. The simpliest case is double-parton
    collisions (two independent hard-scattering occurs in same
    event). The rate such scattering is greatly enhanced in
    nuclear environments. A simple estimation show that ratio
    of double-parton scattering to single hard-scattering is
    $\sigma_2^D/\sigma_1^D \approx 0.5
    (A/10)^{0.5}$~\cite{Strikman:2001gz}.

\item Initial state multiple scattering effects: The rate of
    multiple scattering, where a hard parton undergo one
    hard-scattering plus several soft scattering, is
    significantly enhanced in p+A collisions. Combined with
    steeply falling spectra can shift initial parton to higher
    $p_T$, it can increase the jet pair yield and pedestal
    yield. The fact that near-side jet shape is not modified
    suggests that the multiple scattering happens at the parton
    level before fragmentation.

\item Interaction of initial state radiation with the remaining
    nuclei: In p+p collisions, such radiation simply fragments
    into final state hadrons. But in the d+Au environment, they
    may undergo additional scattering with the nuclei, thus
    further increasing the observed jet pair yield and pedestal
    yield.

\item Cold nuclear jet energy loss:  This effect can increase
    the yield of the soft partons, which in turn leads to more
    pair yield at low $p_T$. However, this scenario also leads
    to a suppression of the high $p_T$ pair yield, which is not
    seen in our data. Thus this effect alone can not explain
    our data.

\item Isospin and EMC effects: It can affect jet pair yield at
    very high $p_T^{sum}$ where the quark jet contribution are
    important. In the kinematic region of our analysis, most
    jet pairs should be dominated by gluon jets, hence these
    effects should not be important.
\end{itemize}
The exact contributions from each of these effects are not clear.
It is possible that several effects conspire to give the observed
modification on the jet correlation and the UE. Future theoretical
efforts should use both single spectra and dihadron correlation
results to better constraint these effects.

In summary, we performed a detailed study of the dihadron
correlation in d+Au collisions over a broad $p_T$ range. A large
Cronin-like enhancement is seen at low pair proxy energy while the
jet shapes show little difference from p+p. This is the same region
where a large modification of the jet properties are also observed
in Au+Au collisions. This observation favors initial state multiple
scattering effects at the partonic level. The jet properties in p+p
are compared to several PYTHIA calculations. The most popular
PYTHIA tunes that work well at Tevtron energy fail to
simultaneously describe the single particle yield, jet pair yield,
and especially the pedestal yield at RHIC energy. The scaling
behavior of the pedestal yield from p+p to d+Au suggests that
underlying event yield in d+Au exceeds a simple sum of one
hard-scattering p+p collision and $N_{coll}-1$ minimum bias p+p
collisions. Theoretical investigation of the centrality dependence
of the underlying event in d+Au can help us to understand the
nature of the multiple parton interaction in p+p and d+Au.



\end{document}